\documentclass{emulateapj}

\usepackage{rotating}

\newcommand{\arcs}{$^{\prime\prime}$}

\shorttitle{The triple system KR Com} \shortauthors{Zasche and Uhl\'a\v{r}}

\begin{document}

\title{\Large \underline{The triple system KR Com} \vskip 7mm}

\author{\large P. Zasche\altaffilmark{1} and R. Uhl\'a\v{r}\altaffilmark{2}  \vskip 3mm}

\affil{ \vskip 3mm \altaffilmark{1} Astronomical Institute, Faculty of Mathematics and Physics,
Charles University of Prague, \\ CZ-180 00 Praha 8, V Hole\v{s}ovi\v{c}k\'ach~2, Czech Republic}

\affil{ \vskip 1mm \altaffilmark{2} Private Observatory, Poho\v{r}\'{\i} 71, 25401 J\'{\i}lov\'e u
Prahy, Czech Republic  \vskip 1mm}


\begin{abstract} \vskip 1mm
\noindent \textbf{\underline{Aims:}}We present the detailed analysis of triple system KR~Com with
different observational techniques  - photometry, interferometry, and period variation.\\
\textbf{\underline{Methods:}} {The use of $BVR$ photometry of the close-contact binary KR~Com,
which is the primary component of a triple system, helps us to better describe the properties of
the components. The interferometric data obtained during the last 30~years sufficiently determine
the visual orbit, but the use of minima timings of KR~Com for the study of period variation
together with the visual orbit is a novel approach in this system.}\\ \textbf{\underline{Results:}}
{Basic physical parameters resulting from the light curve analysis agree well with the previous
results from spectroscopy. The temperatures for the primary and secondary component resulted in
5549 and 6072~K, respectively, and the amount of the third light in all filters is about 1/3 of the
total luminosity. The distant third component revolves around the common barycenter on 11~yr orbit
with a very high eccentricity (0.934) and this movement is also detectable via the period
variation, which is clearly visible in the $O-C$ diagram of times of minima observations. The use
of minima times for the combined analysis helps us to independently determine the distance to the
system ($64.02 \pm 9.42$~pc) and also to confirm the orientation of the orbit in space.}\\
\textbf{\underline{Conclusions: }} {New minima observations and also spectroscopy would be very
profitable, especially during the next periastron passage in the year 2017.}
\end{abstract}


\section{Introduction}

Multiple stellar systems (i.e. of a multiplicity of three and higher) are excellent objects to be
studied. Besides the statistics and their relative frequency among the stars, it is important to
investigate these systems in detail, also because of their stellar evolution, their origin, to test
the influence of the distant components to the close pair, etc.

One of these systems is \object{KR Com} (HD~115955, HIP~65069), which is also A component of the
visual binary A~2166 (WDS~J13202+1747). Its relative brightness is about 7.2~mag in $V$ filter and
its combined spectral type has been classified as F8V, \cite{1981ApJS...45..437A}. The system has
been discovered as a photometrically variable one from the \emph{Hipparcos} data by
\cite{2004A&A...416.1097S}. Its orbital period is about 0.408~days only, but both eclipses are
rather shallow (because of a close bright companion). Although both minima are of similar depths,
the system has been incorrectly classified as a $\beta$~Lyrae one, \cite{1999IBVS.4659....1K}.

The close binary system has been extensively studied spectroscopically by
\cite{2002AJ....124.1738R}. The radial velocity curve of KR~Com has been derived and also an
influence of the third body in the system to the close pair and its properties has been discussed
in subsequent papers by e.g. \cite{2006AJ....132..650D}. A very low value of the mass ratio of this
system has been found, about 0.09 only, which makes KR~Com an exceptional case among the W~UMa-type
binaries. The photometry and light curve solution of the close pair is still missing, which led us
to observe this interesting target.

On the other hand, the close visual companion B (with an angular distance about only 0.1\arcs) was
detected in 1980 by \cite{1983ApJS...51..309M}. The movement around a barycenter is very fast, its
orbital period is about 11~yr and the eccentricity is very high about 0.9. The orbit has been
derived by \cite{1996AJ....111..370H}. Owing to the small angular separation between both
components, any photometric and also spectroscopic observations cannot be done only for KR~Com, and
the influence of the third component has to be considered in the analysis. The eclipsing binary is
the brighter component ($\Delta$m$\sim$ 0.6~mag), and \cite{2002AJ....124.1738R} give the
luminosity ratio $L_3/(L_1+L_2) = 0.56$. They also presented a broadening function, which indicated
a slowly rotating third component, and in \cite{2006AJ....132..650D} the authors provided an
estimate about the third-body mass $M_3 = 1.19$~M$_\odot$, while the individual masses of the close
binary are 1.42 and 0.129~M$_\odot$.

Thanks to the unusually low mass ratio of the eclipsing pair and the presence of the close
interferometric component B we deal with a unique system, which is very useful to be studied in
detail.\\[1mm]

\section{Photometry}

\begin{figure}
 \plotone{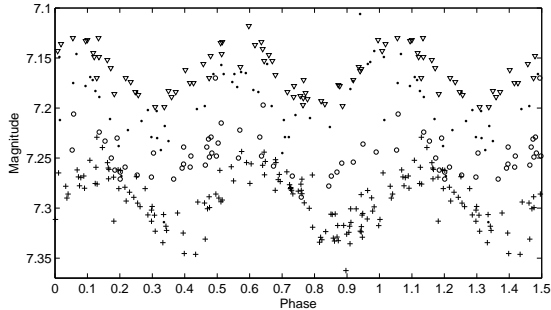}
 \caption{Observations of KR~Com from Hipparcos (plus) and ASAS (circles from 2009, dots from
  2005, and triangles from 2003).} \label{Fig1} \vskip 0.1in ~
\end{figure}

Owing to its relatively high brightness, the system has not been observed photometrically and
studied in detail. There exist the light curves from the \emph{Hipparcos} data ($H_p$ filter), and
from the ASAS survey ($V$ filter), see Fig. \ref{Fig1}. Because of the low photometric amplitude
and high scatter, these can hardly be used for any analysis of the light curve.

Thanks to the observations obtained by ASAS, which cover more than six years, we have found that
there is a long-term photometric decrease of KR~Com. It can be seen in Fig. \ref{Fig1}, where we
present the light curves from from different epochs. As one can see, a steady decrease of its
brightness is on the order of 0.1~mag during the last six years.

\begin{figure}
  \plotone{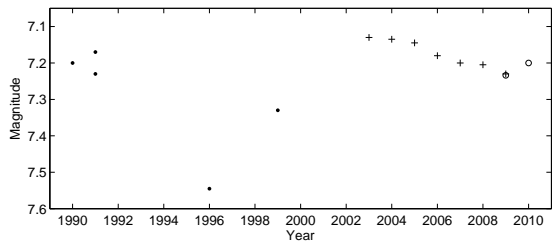}
  \caption{Long-term variation of brightness of KR~Com (in quadratures). Dots represent
  the various catalog data, the plus signs stand for the ASAS data, and the open circles mark
  our observations.}
  \label{Fig1a}
\end{figure}

The nature of this long-term behavior is questionable, as is which component of the system yields
this variation. On the other hand, we have tried to find some historical data in various catalogs,
which contain at least some information about its magnitude in photometric filters. {\sc
VIZIER}\footnote{http://vizier.u-strasbg.fr/viz-bin/VizieR} lists 145 catalogs containing several
data about KR~Com, from which a few usable data points were selected. The most complete are the
measurements in the $V$ filter. In Fig. \ref{Fig1a} we have plotted the data after 1990. The
long-term variation is clearly visible, nevertheless this plot has a few flaws. For some of the
points the photometric filter was not exactly the Johnson's $V$ filer, but rather some "similar"
one. And secondly, for some of the data points it was quite hard to find an exact date of the
observation, only a mean epoch is presented. The most deviating point from 1996 is a typical one.
This particular data point comes from "The PM2000 Bordeaux proper motion catalogue" (Ducourant et
al. 2006), and in its description is mentioned that the filter used is a non-standard one, a
combination of two different filters. Exactly the same problem arises for the second deviating
point from 1999, which comes from the "M2000: An astrometric catalog in the Bordeaux Carte du Ciel"
(Rapaport et al. 2001). However, both these magnitudes use the same combination of filters and the
difference between them during three years is easily visible. A period of this variation could only
be vaguely estimated, but it is probably longer than the period of the third body.

\begin{table}
 \caption{Minima times of KR~Com.}
 \label{Minima}
 \footnotesize
 \centering
 \begin{tabular}{c c c c c}
 \hline\hline
 HJD - 2400000 & Error & Type & Filter & Reference$^{\mathrm{a}}$ \\
 \hline
 48500.2095$\;\;$& 0.0075$\;\;$ & Pri & Hp & HIP  \\
 48500.4200$\;\;$& 0.0129$\;\;$ & Sec & Hp & HIP  \\
 52638.8769$\;\;$& 0.0165$\;\;$ & Sec &  V & ASAS - [1] \\
 52758.41911 & 0.00232 & Sec & V & ASAS \\
 52758.62161 & 0.00301 & Pri & V & ASAS \\
 53058.4796$\;\;$& 0.0002$\;\;$ & Pri &  C & \cite{2005IBVS.5592....1K} \\
 53102.11786 & 0.00255 & Pri & V & ASAS \\
 53150.2695$\;\;$& 0.0005$\;\;$ & Pri &  C & \cite{2005IBVS.5592....1K} \\
 53475.0061$\;\;$& 0.0004$\;\;$ & Pri &  V & \cite{VSOLB2006} \\
 53487.86332 & 0.00518 & Sec & V & ASAS \\
 53488.05344 & 0.00767 & Pri & V & ASAS \\
 53829.11608 & 0.00468 & Sec & V & ASAS \\
 54225.87197 & 0.00849 & Sec & V & ASAS \\
 54226.07038 & 0.00326 & Sec & V & ASAS \\
 54233.4095$\;\;$& 0.0009$\;\;$ & Pri &  C & [2] \\
 54580.40196 & 0.00273 & Sec & V & ASAS \\
 54613.0406$\;\;$& 0.0002$\;\;$ & Sec & Ic & \cite{VSOLB2009} \\
 54921.47502 & 0.00150 & Sec  &  B & this paper \\
 54921.47435 & 0.00246 & Sec  &  V & this paper \\
 54921.47702 & 0.00178 & Sec  &  R & this paper \\
 54921.47534 & 0.00051 & Sec  &  I & this paper \\
 54924.53515 & 0.00043 & Pri  &  B & this paper \\
 54924.53553 & 0.00048 & Pri  &  V & this paper \\
 54924.53559 & 0.00227 & Pri  &  R & this paper \\
 54924.53549 & 0.00031 & Pri  &  I & this paper \\
 54940.44566 & 0.00113 & Pri  &  R & this paper \\
 54971.44977 & 0.00098 & Pri  &  B & this paper \\
 54971.45562 & 0.00085 & Pri  &  V & this paper \\
 54971.45419 & 0.00052 & Pri  &  R & this paper \\
 54944.52261 & 0.00457 & Pri & V & ASAS \\
 54944.32228 & 0.00565 & Sec & V & ASAS \\
 55259.67912 & 0.00319 & Sec  &  R & this paper \\
 55274.37115 & 0.00175 & Sec  &  B & this paper \\
 55274.36049 & 0.00135 & Sec  &  I & this paper \\
 55274.36870 & 0.00543 & Sec  &  R & this paper \\
 55274.35921 & 0.00329 & Sec  &  V & this paper \\
 55274.56764 & 0.00335 & Pri  &  B & this paper \\
 55274.56959 & 0.00143 & Pri  &  I & this paper \\
 55274.56684 & 0.00293 & Pri  &  R & this paper \\
 55274.56834 & 0.00264 & Pri  &  V & this paper \\
 55280.49301 & 0.00093 & Sec  &  B & this paper \\
 55280.49220 & 0.00069 & Sec  &  R & this paper \\
 55280.48942 & 0.00109 & Sec  &  V & this paper \\
 55281.50792 & 0.00089 & Pri  &  B & this paper \\
 55281.50389 & 0.00085 & Pri  &  R & this paper \\
 55281.50817 & 0.00050 & Pri  &  V & this paper \\
 55293.53729 & 0.00061 & Sec  &  B & this paper \\
 55293.53830 & 0.00157 & Sec  &  R & this paper \\
 55293.53559 & 0.00099 & Sec  &  V & this paper \\
 55294.56261 & 0.00110 & Pri  &  B & this paper \\
 55294.56418 & 0.00157 & Pri  &  R & this paper \\
 55294.56473 & 0.00120 & Pri  &  V & this paper \\
 55357.38912 & 0.00235 & Pri  &  I & this paper \\
 55358.41029 & 0.00257 & Sec  &  I & this paper \\
  \hline
\end{tabular}
\begin{list}{}{}
\item[$^{\mathrm{a}}$] HIP - \emph{Hipparcos} observations, ASAS - data from the ASAS survey, [1] -
see http://var.astro.cz/ocgate, [2] - unpublished yet, see
 http://eclipsingbinary.web.fc2.com/vsnetmin.htm
\end{list}
\end{table}

One can speculate about the magnetic activity cycles in the system. Because of the spectral type
(F8V), these can play a role, see \cite{Hall1989}. These cycles could slightly modulate the orbital
period of the close pair. Moreover, accepting this hypothesis, according to \cite{Applegate1992}
there is also a modulation of the luminosity of the star on the order of $\delta L / L \sim 0.1$.
If we propose that the most luminous component (the primary one) undergoes this variation, a
variability of the brightness of the whole system results in only about 0.05~mag, which is much
lower than observed since 1990. On the other hand, removing these two questionable data points
discussed above, one can obtain a variation with an amplitude of about only 0.1~mag, which could be
described by this mechanism. Therefore, the nature of this variation still remains an open
question.

We have observed the system during the seasons 2009 and 2010. In total there are 17 nights of
observations, but for the light-curve analysis we used only 7 nights of observations obtained from
March 2010 to April 2010 and carried out with the same telescope and detector at the private
observatory by one of the authors (RU). Owing to high brightness of the target, the refractor with
a diameter of only 75~mm was used, equipped with the G2/KAF 0402ME CCD camera and standard $B$,
$V$, and $R$ filters according to specification by Bessell (1990). All the measurements were
processed by the software {\sc C-Munipack}\footnote{see
$\mathrm{http://c-munipack.sourceforge.net/}$}, which is based on aperture photometry and using the
standard DaoPhot routines (Tody 1993).

The observations were transformed into the standard system using the well-known transformation
equations\footnote{see $\mathrm{ http://brucegary.net/AllSky/x.htm}$}. Star \object{HD 115981} was
used as a comparison, while the check star to control the non-variability of the two was star
\object{HD 116206}. The atmospheric extinction has been neglected because the stars are very close
to each other (11~arcmin), their spectral types are also similar, and the observations have never
been obtained below 25~deg above horizon. We used this approach because our observations have an
accuracy not better than 0.01~mag (for the scatter in individual filters see below), therefore any
additional correction on the order of 0.001~mag is practically useless for the transformation. The
individual exposure times range from 25 to 140~seconds.

The rest of observations were used only to determine the minima times for a period analysis. These
new ones as well as the already published ones are given in Table \ref{Minima}, where the type of
minima refers to the following ephemeris: $2454924.538 + 0.4079711 \cdot E.$ For all these
observations, the Kwee - van~Woerden method (hereafter KW, Kwee \& van~Woerden 1956) was used for
determining the time of minimum. This method is suitable for symmetric minima (which is the case
for KR~Com), because its principle is based on comparing the ascending and descending branch of the
minima. We do believe that our error estimates are much more reliable than those already published
because first, we did a detailed analysis of the KW result for each minimum, which means that we
used different data sets (neglecting some of the observed points) and compared these results. And
secondly, we also used a polynomial fitting to determine the time of minimum and compared this
result with the KW one. The error of the particular minimum light was computed as a maximum
difference between all the different results from different methods from the mean value. All minima
times given in Table \ref{Minima} are heliocentric ones. Eleven new minima times were also derived
from the data of the ASAS survey (Pojmanski 2002).\\[1mm]

\section{The light curve analysis}

For the light-curve analysis we used the program {\sc PHOEBE} (Pr{\v s}a \& Zwitter 2005), which is
based on the Wilson-Devinney algorithm (Wilson \& Devinney 1971). The model atmospheres made by
Kurucz (1996) were used. The derived quantities are: the individual temperatures $T_1$ and $T_2$,
the inclination $i$, the luminosities $L_i$, the gravity darkening coefficients $g_i$, the limb
darkening coefficients $x_i$, the albedo coefficients $A_i$, and the synchronicity parameters
$F_i$. The limb darkening was approximated via linear cosine law, and the values of $x_1$ and $x_2$
were interpolated from the van~Hamme's tables, \cite{vanHamme1993}.

\begin{table}
 \caption{Light curve parameters of KR~Com.}
 \label{LCparam}
 \scriptsize
 \centering
 \begin{tabular}{c c | c c}
 \hline\hline
 Parameter & Value & Parameter & Value \\
 \hline
 $T_1$ [K]  & 5549 $\pm$ 244    &    $L_1$ (B) [\%] & 50.67 $\pm$ 0.20 \\
 $T_2$ [K]  & 6072 $\pm$ 270    &        $L_2$ (B) [\%] & 17.75 $\pm$ 0.19 \\
 $q$ ($={M_2}/{M_1}$) & 0.091 (fixed) &  $L_3$ (B) [\%] & 31.58 $\pm$ 0.20 \\
 $e$        &  0 (fixed)        &        $L_1$ (V) [\%] & 47.63 $\pm$ 0.24 \\
 $i$  [deg] & 52.14 $\pm$ 0.46  &        $L_2$ (V) [\%] & 18.98 $\pm$ 0.23 \\
 $x_1$ (B)  & 0.774             &        $L_3$ (V) [\%] & 33.39 $\pm$ 0.24 \\
 $x_2$ (B)  & 0.699             &        $L_1$ (R) [\%] & 47.05 $\pm$ 0.72 \\
 $x_1$ (V)  & 0.635             &        $L_2$ (R) [\%] & 19.04 $\pm$ 0.67 \\
 $x_2$ (V)  & 0.565             &        $L_3$ (R) [\%] & 33.90 $\pm$ 0.72 \\ \cline{3-4}
 $x_1$ (R)  & 0.548             &  \multicolumn{2}{c}{Derived quantities:}  \\
 $x_2$ (R)  & 0.485             &  $R_1$ [R$_\odot$] & 1.33 $\pm$ 0.04 \\
 $g_1$      & 0.323 $\pm$ 0.014 &  $R_2$ [R$_\odot$] & 0.49 $\pm$ 0.02 \\
 $g_2$      & 0.351 $\pm$ 0.021 &  $M_{\mathrm{bol,1}}$ [mag] & 4.34 $\pm$ 0.18 \\
 $A_1$      & 0.499 $\pm$ 0.027 &  $M_{\mathrm{bol,2}}$ [mag] & 6.13 $\pm$ 0.24 \\ \cline{3-4}
 $A_2$      & 0.484 $\pm$ 0.015 &  $\sigma$ [mag] (B) &  0.00745 \\
 $F_1$      & 1.690 $\pm$ 0.066 &  $\sigma$ [mag] (V) &  0.01583 \\
 $F_2$      & 1.647 $\pm$ 0.148 &  $\sigma$ [mag] (R) &  0.06737 \\ \hline \hline
 \end{tabular}
\end{table}

\begin{figure}[b]
  \centering
  \plotone{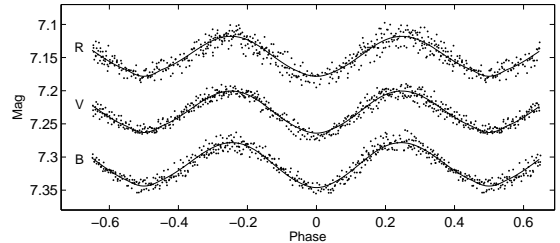}
  \caption{Our new observations of the light curves of KR~Com in three filters. The $B$ and $R$ curves
  are shifted for better clarity. The theoretical fits and parameters are described in the text
  and Table \ref{LCparam}.}
  \label{Fig2}
\end{figure}

Because the system has been analyzed spectroscopically in detail by \cite{2002AJ....124.1738R}, we
used the mass ratio value $q=M_2 / M_1 = 0.091$ derived from spectroscopy for the analysis of our
$B$, $V$, and $R$ photometry. The other relevant parameters of the light curve were computed (with
fixed circular orbit) and the results are presented in Table \ref{LCparam}. The final fits and the
photometric data are plotted in Fig. \ref{Fig2}. The scatter in the $B$ and $V$ filters is about
0.010-0.015~mag, in the $R$ about 0.025~mag, and the numbers of data points are 451, 468, and 474,
for $B$, $V$, and $R$, respectively. The individual observational errors of the measurements were
about 5-10 times lower than a typical peak-to-peak scatter of the measurements in the particular
filter. The same applies for the measurement errors for the comparison and check stars. Their
respective errors are also on the order of 0.003~mag only, but these values represent merely the
mathematical results of errors from the photometry of the stars. More plausible physical errors are
much above these values, about 0.01~mag. This is actually a very similar situation as for the
minimum estimate and the error of the KW method as discussed in the previous section. No additional
short-time variation has been found in the data. The $BVR$ photometry used for the analysis is also
available as an online table.

The errors of the individual parameters were derived from the covariance matrix, following a
standard procedure of error estimates, see e.g. "Numerical recipes. The art of scientific
computing", \cite{1986nras.book.....P}.

From the light-curve analysis we see that the two components are in contact (A-type contact
system), but their individual temperatures differ about more than 500~K. On the other hand, the
difference between the temperatures of $523 \pm 364$~K is only marginally significant. We tried to
find an appropriate solution with equal temperatures (which resulted in a value of about 6100~K),
but this solution resulted in $\chi^2$ about 13\% worse than the solution presented in Table
\ref{LCparam}. Another approach was to fix the temperature of the primary to $T_1 = 6100~$K and to
fit only the secondary temperature. This resulted in fit with $\chi^2$ about 9\% worse than our
final solution. Both alternatives have been tested via an F-test to gauge how liable they are to
introduce another free parameter (temperature) to the model. We concluded that introducing this new
parameter is significant at a level about 86\% and 84.5\% for these two different approaches.

Our resulting values of parameters from the light curve fit can be compared with the estimates
published in \cite{2002AJ....124.1738R}. A value of the third light (33\% of total light) agrees
well with their estimate (about 56\% of $L_1 + L_2$). The individual components are probably of F+G
spectral types for the primary and secondary, respectively. The secondary component has apparently
undergone a mass transfer to the primary.

We were able to partly reveal the nature of the third body in the system via its temperature. This
can be estimated thanks to the individual luminosities in the filters, from which one can derive
the magnitudes in the particular filters. These values resulted in $B=8.96$~mag, $V=8.39$~mag, and
$R=8.05$~mag. As an independent comparison one can also check the value from the Washington Double
Star Catalog (WDS)\footnote{http://ad.usno.navy.mil/wds/}, where the $V$~magnitudes 7.78 and
8.38~mag are given for the primary and secondary component of the visual pair, which excellently
agrees with our value of 8.39~mag. From our estimates of the $BVR$ magnitudes, the photometric
indices $B-V=0.57$~mag and $V-R=0.34$~mag can be compared with the color-temperature relations,
e.g. by \cite{2000AJ....119.1448H}. This gives temperature estimate of the third component of $5900
\pm 200$~K, which indicates a spectral type between F8 and G6. On the other hand, if we try to
derive also the radius of this component from the relation between the radii, luminosities and
temperatures $R_3 = \sqrt{L_3/L_1 \cdot (T_1/T_3)^4 }\cdot R_1$, we obtain values from 0.93 to
0.99~R$_\odot$, which indicates a later spectral type, about G8 - K1 (Harmanec 1988).\\[1mm]

\section{Combined solution for the visual orbit and period variations}

As we mentioned above, the A-B system revolves around a barycenter with a period about 11~yr.
Because this orbit has been derived about 15~years ago and many new observations are available
since then, there is a need for new up-to-date solution for this orbit.

We used all available interferometric observations for the analysis of the pair collected in the
"Fourth Catalog of Interferometric Measurements of Binary
Stars\footnote{http://ad.usno.navy.mil/wds/int4.html}", \cite{2001AJ....122.3480H}. There are 25
usable measurements (see Table \ref{interferometry}). In some of these observations the position
angle $\theta$ has to be changed by 180$^\circ$ (i.e. quadrant change, interchange of the
components), the values presented in Table \ref{interferometry} are the corrected ones.

\begin{table}
 \caption{Interferometric observations of A-B pair.}
 \label{interferometry}
 \small
 \centering
 \begin{tabular}{c c c c} 
 \hline\hline
 Date & $\theta$ & $\rho$   & References \\
      &  [deg]   & [arcsec] &  \\
 \hline
 1980.1566 &   0.9 & 0.145 & \cite{1983ApJS...51..309M} \\
 1980.4817 &   2.0 & 0.142 & \cite{1983ApJS...51..309M} \\
 1983.0510 &  10.0 & 0.117 & \cite{1987AJ.....93..688M} \\
 1983.0701 &  12.2 & 0.106 & \cite{1987AJ.....93..688M} \\
 1984.0558 &  22.6 & 0.059 & \cite{1996AJ....111..370H} \\
 1986.4067 & 327.7 & 0.058 & \cite{1989AJ.....97..510M} \\
 1987.1194 & 345.5 & 0.076 & \cite{1997AJ....114.1623F} \\
 1987.2642 & 340.2 & 0.083 & \cite{1989AJ.....97..510M} \\
 1987.2859 & 343.0 & 0.089 & \cite{1997AJ....114.1623F} \\
 1987.2886 & 348.8 & 0.089 & \cite{1997AJ....114.1623F} \\
 1987.2914 & 346.8 & 0.089 & \cite{1997AJ....114.1623F} \\
 1988.1022 & 345.5 & 0.097 & \cite{1997AJ....114.1623F} \\
 1989.2300 & 353.6 & 0.117 & \cite{1990AJ.....99..965M} \\
 1990.2081 & 357.9 & 0.132 & \cite{1994AAS..105..503B} \\
 1990.2621 & 356.0 & 0.136 & \cite{1992AJ....104..810H} \\
 1990.2759 & 356.0 & 0.135 & \cite{1992AJ....104..810H} \\
 1991.25   & 357.0 & 0.136 & \cite{HIP} \\
 1991.3187 &   0.7 & 0.140 & \cite{1994AJ....108.2299H} \\
 1992.3098 &   4.3 & 0.138 & \cite{1994AJ....108.2299H} \\
 1992.3127 &   4.2 & 0.139 & \cite{1994AJ....108.2299H} \\
 1993.1971 &   7.7 & 0.128 & \cite{1994AJ....108.2299H} \\
 2002.3224 & 359.8 & 0.140 & \cite{2008AJ....136..312H} \\
 2002.3224 &   0.0 & 0.140 & \cite{2008AJ....136..312H} \\
 2002.3224 &   0.0 & 0.139 & \cite{2008AJ....136..312H} \\
 2006.1917 &  17.8 & 0.077 & \cite{2009AJ....137.3358M} \\
 \hline
 \end{tabular}
\end{table}

A movement of the contact binary around the barycenter with the distant component has to produce a
periodic variation of orbital period of KR~Com, a well-known "light-time effect" (hereafter LITE),
see \cite{Irwin1959}. The amplitude of the LITE depends on the inclination of the 11~yr orbit, the
individual masses and also on the distance of KR~Com. The distance from \emph{Hipparcos} was
originally derived as $76.51 \pm 5.46$~pc, i.e. parallax $\pi = 13.07 \pm 0.87$~mas (Perryman \&
ESA 1997), but the more recent value is $83.97 \pm 5.32$~pc, i.e. parallax $\pi = 11.91 \pm
0.71$~mas, \cite{2007A&A...474..653V}. Nevertheless, the value of the parallax from the
\emph{Hipparcos} satellite could also be influenced by the movement on the long orbit. During the
run of the satellite, the change of the position angle of the distant component was only about
15$^\circ$, because in that time the B component was near apastron on its orbit. On the other hand,
if one compares the error of its parallax with the other stars from the \emph{Hipparcos} catalog
with the similar parallaxes, the error is apparently not deviating significantly.

\begin{table}
 \caption{Final parameters of the long orbit.}
 \label{Final}
 \small
 \centering
 \begin{tabular}{c c c c}
 \hline\hline
 Parameter & Value \\
 \hline
 $HJD_0$       & 2453058.4640 $\pm$ 0.0021 \\
 $P$  [day]    & 0.40797003 $\pm$ 0.00000239 \\
 $p_3$ [day]   & 4011.0 $\pm$ 98.4 \\
 $p_3$ [yr]    & 10.98 $\pm$ 0.27 \\
 $T_0$         & 2442055.1 $\pm$ 118.0 \\
 $a$ [arcsec]  & 0.1131 $\pm$ 0.0189  \\
 $\omega_3$ [deg] & 301.8 $\pm$ 3.8 \\
 $e_3$          & 0.9340 $\pm$ 0.0239 \\
 $i$ [deg]      & 67.68 $\pm$ 2.03 \\
 $\Omega$ [deg] & 210.03 $\pm$ 1.25 \\
 $A$  [day]     & 0.0171 $\pm$ 0.0019 \\ \hline
 $f(M_3)$ [M$_\odot$] & 0.3262 $\pm$ 0.0896 \\
 $M_3$ [M$_\odot$]  & 1.598 $\pm$ 0.423 \\
 $a_{12}$ [AU]  & 3.677 $\pm$ 0.821 \\
 $a_3$   [AU]   & 3.563 $\pm$ 1.027 \\
 $\pi$ [mas]    & 15.62 $\pm$ 1.80 \\
 \hline
 \end{tabular}
\end{table}

\begin{figure}
  \centering
  \plotone{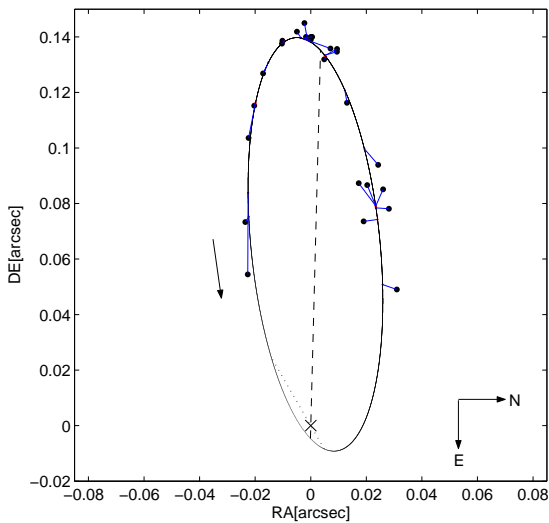}
  \caption{Orbit of A-B pair on a plane of the sky. The dashed line represents the line of the
  apsides, while the dotted one stands for the line of the nodes. The cross indicates the position of
  the eclipsing binary.}
  \label{Fig3}
\end{figure}

For an analysis of the period changes of KR~Com we collected all available minima observations. The
very first ones are from \emph{Hipparcos} data (recalculated primary minimum and derived also a
secondary one), while the most recent ones are our new observations of 2010. All these heliocentric
minima times are given in Table \ref{Minima}. We found that all the published minima times have
their respective errors strongly underestimated (10 times or even more).

We used a similar approach as described in \cite{ZascheWolf} for analyzing the interferometric data
together with the times of minima. The only difference was a calculation of both amplitudes - the
semimajor axis of the visual orbit and the semiamplitude of LITE in the $O-C$ diagram. The main
advantage of this approach is that we can independently calculate the distance (or parallax) to the
system and compare it to the \emph{Hipparcos} one. Therefore, the set of parameters to be computed
is the following: $HJD_0, P, p_3, T_0, i, a, \omega, \Omega, e_3, A,$ where $a$ denotes the
semimajor axis in arcseconds, while $A$ stands for the semiamplitude of LITE. We used the
least-squares method and the simplex algorithm (see e.g. Kallrath \& Linnell 1987). One can also
discuss the use of quadratic ephemeris because of some mass transfer between the components, but
this was not used in our analysis due to insufficient coverage of the $O-C$ diagram with data
points in longer time scales. The two minima times based on \emph{Hipparcos} data were only roughly
estimated and therefore their use for the analysis is questionable.

\begin{figure}
  \centering
  \plotone{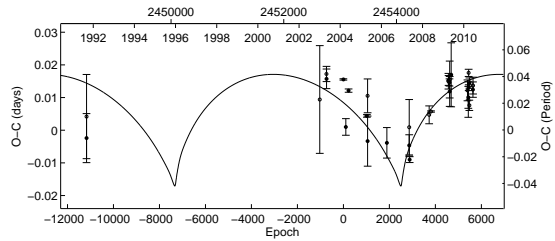}
  \caption{$O-C$ diagram of KR~Com. The primary minima are plotted as dots, the secondary ones as open
  circles. It is obvious that the respective errors of the published minima times should be larger
  than presented.}
  \label{Fig4}
\end{figure}

\begin{figure}
  \centering
  \plotone{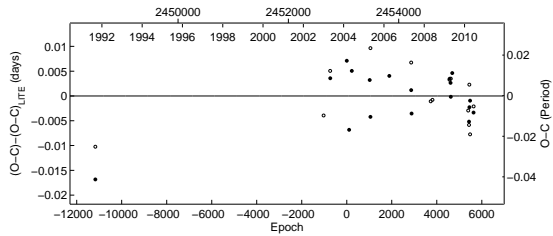}
  \caption{$O-C$ diagram of residuals of KR~Com.}
  \label{Fig5}
\end{figure}

The analysis results in a set of parameters presented in Table \ref{Final}, where we list all
computed values and some of the derived quantities. The final fits are presented in Figs.
\ref{Fig3} and \ref{Fig4}. For plotting the $O-C$ diagram, the individual minima in different
filters were averaged to one point for better clarity. Figure \ref{Fig5} shows the residuals after
subtraction of the fit. As one can see, some long-term variation cannot be ruled out with the
current data. This could be caused e.g. by a mass transfer between the components, but only future
observations can confirm or rule out this possibility. Another potential explanation are magnetic
cycles with a period of about 20~years (Applegate 1992), which cause not only a brightness
variation, but also this period variation. A combination of both mass transfer and magnetic cycles
is also possible.

\begin{figure}
  \centering
  \plotone{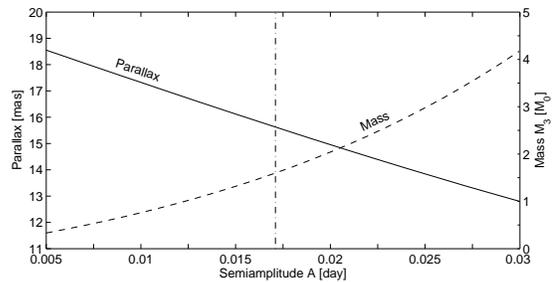}
  \caption{Diagram of semiamplitude of LITE versus mass of the third body and parallax of the system with the
  fixed semimajor axis of the visual orbit. The dash-dotted line represents the value of $A$ from our final solution.}
  \label{Fig6}
\end{figure}

The distance to the system was derived from the fit, resulting in $d = 64.02 \pm 9.42$~pc, closer
than from the \emph{Hipparcos} data. However, this value is only roughly estimated because it is
derived from comparing the values of $A$ and $a$, but the detection of the LITE variation and
deriving its amplitude is still problematic, and the 11~yr period is not yet sufficiently covered
by data. Due to this reason, and due to the relatively large difference between our derived
parallax and the value from \emph{Hipparcos}, we introduced Fig. \ref{Fig6}. In this figure we
fixed the semimajor axis of the visual orbit (this value is much better defined than the amplitude
of LITE) and plotted the values of the parallax $\pi$ and mass of the third component $M_3$ with
respect to the amplitude $A$ of LITE. As one can see, our resulting value of $A = 0.0171$~days
yields the values of $M_3$ and $\pi$ given in Table \ref{Final}. If we consider the range of
potential values of $A$ according to its error, the values of $\pi$ are still far from the value
derived by \emph{Hipparcos}. Moreover, the mass of the third body with a decreasing value of
parallax is growing rapidly. On the other hand, if we presume that the mass of the third component
is lower, about e.g. 1.1~M$_\odot$, then the amplitude of LITE is only about 0.013~days (which
cannot reliably describe the fit in the $O-C$ diagram) and the parallax results in approximately
16.5~mas, which is even more distant than the \emph{Hipparcos} value.

The other quantity, which can be compared with the previous results, is e.g. the mass of the
distant component $M_3$. \cite{2006AJ....132..650D} presented a value about 1.19~M$_\odot$, while
our estimation is a bit higher, about 1.60~M$_\odot$. This value has been derived from the
assumption that the masses of the individual components are $M_1 = 1.42$~M$_\odot$ and $M_2 =
0.129$~M$_\odot$, which have been taken as constant input values. The resulting value of $M_3$ is
surprising, especially if we take into consideration that the luminosities in all filters resulted
in slightly lower values ($B$: 46\%, $V$: 50\%, $R$: 51\% of $L_1+L_2$) than assumed by
\cite{2006AJ....132..650D}. If the mass of the third body is higher and the luminosity lower, this
could indicate that this component is also a binary, or the star is underluminous than one would
expect for a star of the main sequence.\\[1mm]

\section{Discussion and conclusions}

The triple system KR~Com revealed some of its basic physical properties for the first time. The
light curve with shallow minima (due to a bright third component) shows that the system has a
relatively low inclination of about only 52$^\circ$. That we know the inclinations of both orbits
is one aspect of KR~Com system's uniqueness.

On one hand, a difference between the inclination of the close pair and the inclination of the
orbit of the third distant component ($i_3 = 67.8^\circ$) is still relatively small, therefore one
could speculate about a common origin of all three components (see e.g. Zakirov 2008). On the other
hand, a presence of two orbits which are not strictly coplanar is a necessary condition for a
so-called Kozai cycles (Kozai 1962), which could play a role in the formation of this system. As
mentioned by \cite{2006AJ....132..650D}, the typical product of the Kozai mechanism is a close pair
and a distant component on a non-coplanar orbit, which is the case here.

The other computable quantity in this triple system is its nodal period, or the precession of the
orbits, which can change the inclination of the close eclipsing pair and therefore also the depths
of the minima, see e.g. \cite{2005Ap&SS.296..113M}. Regrettably, the nodal period here results in a
value of about 100000~yr.

Today there are only 34 eclipsing binary systems known among the visual double stars that have
known visual orbits. In all these systems one should expect a period variation of times of minima
observed for these systems, but surprisingly, the LITE that agrees with the visual orbit has been
detected in only seven systems (\object{i Boo}, \object{VW Cep}, \object{$\zeta$ Phe}, \object{V819
Her}, \object{V772 Her}, \object{QS Aql}, \object{V505 Sgr}). The system KR~Com seems to be another
example. The visual orbit with a period about 11~yr is well-covered by interferometric data, and
the combined solution of visual orbit and $O-C$ diagram of minima times is also dominated by the
visual orbit. Therefore, the very first minima estimates based on \emph{Hipparcos} data with their
large errors were almost useless for the analysis. The other interesting fact is that the already
published minima show much larger residuals than their published errors.

However, the combined fit brings new light into the system and confirms for example the $\omega$
and $\Omega$ values. Because the same fit to the interferometric data could be realized by changing
both angles by 180$^\circ$ (orientation of the orbit towards the observer), the LITE fit helps us
to confirm the orientation of the orbit in the space without any need for spectroscopy.
Nevertheless, the spectroscopic data and new times of minima would be also very profitable,
especially in the upcoming periastron passage, which will occur in 2017.

\begin{acknowledgements}
This investigation was supported by the Research Programme MSM0021620860 of the Czech Ministry of
Education. We thank P.Svoboda for providing us his photometric CCD observations. We also do thank
the ASAS team for making all of the observations easily public available. An anonymous referee is
also acknowledged for his or her helpful and critical suggestions, which greatly improved the
paper. This research has made use of the SIMBAD and VIZIER databases, operated at CDS, Strasbourg,
France and of NASA's Astrophysics Data System Bibliographic Services.
\end{acknowledgements}

\end{document}